\input{aipcheck}
\documentclass[,draft] {aipproc}
\usepackage{epsfig,epsf,amssymb,amsmath}
\layoutstyle{6x9}
\begin{document}
\title{Anomalous Internal Pair Conversion\\
Signaling Elusive Light Neutral Particles}
\author{Fokke de Boer}{
  address={Nationaal Instituut voor Kernfysica en Hoge-Energie Fysica.\\ NL-1098 SJ Amsterdam, The Netherlands}
}

\date{}

\begin{abstract}
In this paper we report on a systematic search 
for a neutral boson in the mass range between 5 and 15 MeV/c$^{2}$
in the decay of highly excited nuclei. 
Its signature is found a deviation in the angular
correlation of the $e^+e^-$ pairs from 
conventional internal pair conversion (IPC) resulting from 
of its two-body decay kinematics.
With an $e^{+}e^{-}$ pair-spectrometer, 
a number of transitions has been investigated in the 
 ${\alpha}$--nuclei $^{8}$Be, 
$^{12}$C
and $^{16}$O, following light ion induced reactions 
at low bombarding energies, first  
at IKF in Frankfurt and during the last years at ATOMKI in Debrecen.  
Startlingly, in all isoscalar transitions excess $e^{+}e^{-}$ pairs are found 
at large angles with branching ratios with respect to the total yield
ranging from 10$^{-2}$ to 10$^{-6}$.  
If these deviations are all related to the two-body decay of an $X$-boson, 
this observation implies plural $X$-bosons. 
An analysis of all angular spectra with a boson 
search program, yields a pandemonium of more than ten 
candidate bosons.  

\end{abstract}

\maketitle


\subsection{Introduction}

Elusive neutral $X$-bosons can be emitted in a nuclear 
transition within the constraints of spin-parity, isospin and energy-momentum 
conservation. 
Superallowed ({\em L=0)} emission is expected for a  {\em pseudoscalar}   
 $0^{-}$ particle in a $0^{-} {\rightarrow} 0^{+}$ M0 transition, 
a  {\em scalar} particle   
$0^{+}$ in an $0^{+} {\rightarrow} 0^{+}$ E0, an  {\em  axial vector} $ 1^{+}$
in an M1 and a {\em vector} $1^{-}$ particle in an E1 transition. Allowed 
({\em L=1}) emission occurs for  $0^{+}$  and  $0^{-}$  particles in E1 
and M1 respectively,
and for $1^{+}$ and $1^{-}$ in E1, M1 but also in E2 and M2 transitions.  
The signature of $X$-boson emission is expected to be 
the two-body decay into $e^+e^-$ pairs 
superposed on conventional 
internal pair conversion (IPC) \cite{rose}, but with a marked difference
in the distribution of the $e^{+}e^{-}$ opening angle.
 
Since some time a search is carried out for 
 such a neutral particle in 
high-energy transitions in self-conjugate ${\alpha}$-nuclei. 
Since 1990, experiments have been performed at the 'small' Van de Graaff 
accelerator of the Institut f\"{u}r Kernphysik in Frankfurt, Germany and   
since the year 2000 at the AVF cyclotron and the Van de Graaff accelerator at 
ATOMKI in Debrecen, Hungary. In all experiments the same compact 
spectrometer was used, consisting of eight E-${\Delta}$E detectors, allowing a 
good compromise between angular granularity, stopping power, range and pair-
 efficiency \cite{stiebing}. 

A signal which could originate from a boson with an invariant mass of 
9 MeV/c$^{2}$ has been observed in the M1 transitions depopulating 
the 17.64 MeV, $1^{+}$ level in $^{8}$Be to the ground- and first excited state. 
In Table 1 of refs. \cite{fokke2,fokke3,fokke4}, the results are collected 
which are relevant to this anomaly for the ${\alpha}$-nuclei $^{4}$He 
to $^{20}$Ne, like the  branching ratio $B_X$, the width ${\Gamma}_X$, 
the coupling strength ${\alpha}_X$ and the $m_{X}$ for the different 
transitions.  
It appears that isoscalar magnetic transitions 
and a possible $0^{-} {\rightarrow} 0^{+}$ transition 
exhibit an excess of $e^{+}e^{-}$ pairs at large opening angles,
whereas no deviations occur in isovector E1 and M1 transitions.
The deviations in 
isoscalar M1 transitions indicate an isoscalar character 
for such a boson with spin-parity 0$^{-}$ or 1$^{+}$, and 
coupling strength ${\alpha}_X$ proportional to the 
isoscalar strength of the M1 transition.  
The sensitivity of these $X$-boson 
searches---expressed in the ratio of ${\Gamma}_X$ to ${\Gamma}_{\gamma}$--varies
between 10$^{-6}$ and 10$^{-2}$. 

In order to test this $X$-boson scenario, we started a dedicated search
of $e^{+}e^{-}$ angular correlations in isoscalar magnetic transitions
with inelastic proton scattering experiments
on $^{12}$C and $^{16}$O, the $(p,{\alpha})$ reaction on $^{19}$F 
and the ($^{3}He,pe^{+}e^{-}$) reaction on $^{10}$B and $^{14}$N targets.
The first aim was to check the signal observed \cite{fokke4}  
for the isoscalar M1 ground state transition from the  
T=0 level at 12.71 MeV in $^{12}$C
at a high level of significance and the second aim to look 
for the 9 MeV/c$^{2}$ boson in  
the 'forbidden' $0^-{\rightarrow}0^+$ (M0) transition at 10.96 MeV in $^{16}$O. 
The mere observation of $e^{+}e^{-}$ decay to the ground state 
imposes a straightforward bound on an $X$-boson emission width.

\subsection{Experiments}

Experiments are carried out
at ATOMKI, in Debrecen, Hungary 
at the cyclotron and at the Van de Graaff accelerator.  
The first four experiments (DEB1-DEB4) were dedicated to the  
$^{12}$C$(p,p{'}e^{+}e^{-})^{12}$C$^*$ and the  
 the $^{16}$O$(p,p^{'}e^{+}e^{-})^{16}$O$^*$ reaction at 16.5 MeV at the
cyclotron using only the six 'small' detectors.  
For calibration the $^{19}$F$(p,{\alpha}e^{+}e^{-})^{16}$O$^*$ 
reaction was used at 3.5 and 5.5 MeV. 
In the DEB5-run, the $^{14}$N$(^{3}He,pe^{+}e^{-})^{16}$O$^*$ 
reaction was studied at 2.3 MeV together with the $^{19}$F$(p,{\alpha}e^{+}e^{-})^{16}$O$^*$ reaction at 3.5 MeV using the complete eight detector set.
The DEB7 experiment was performed in an 'asymmetric' 
arrangement with five telescopes (two large and three small ones) and 
an additional detector for proton detection.   
The $^{10}$B$(^{3}He,pe^{+}e^{-})^{12}$C$^*$ reaction was measured
followed by the $^{14}$N$(^{3}He,pe^{+}e^{-})^{16}$O$^*$ reaction 
for comparison. These reactions were found to favor the population
of unnatural parity states \cite{bromley,almquist}. 
In this experiment a thin proton detector with an 11\%
efficiency was implemented downstream. In coincidence 
with the populating proton group, $e^{+}e^{-}$ pairs from the different 
levels could be distinguished. 

In comparison with the $p$-capture spectra obtained at IKF where
the decay of only one level is selected by the choice of the bombarding energy,
the  $(p,{\alpha}$),
$(^{3}He,p)$ and $(p,p')$ reactions have several entrance channels in the
final nuclei. The resulting $e^{+}e^{-}$ sum-spectra are composite,
but nevertheless different sum-peaks can be distinguished. 
We clearly observe deviations in the angular correlation spectra 
of transitions with energies below 9 MeV, the aimed boson mass,
which energetically cannot decay by this particle. 
In the particle scenario with 
($m_{X}{\sim}E{\cdot}sin({\omega}/2))$, these 
deviations have to be explained by the emission of different particles with lower mass. With the observation of several deviations, a number of new
particles has to be postulated. This also opens the exciting possibility
of multiple boson emission in one transition.

Analysis of the presumably composite spectra with the number of bosons,
their mass and strength as free parameter is not trivial and time-consuming. 
We have solved this problem in an indirect approach. We converted the 
angular spectrum in a 'quasi'-mass spectrum by comparing
the ratio of experimental data-points, divided by the simulated IPC
distribution, with the ratio obtained by simulation of e.g. $10^{8}$ bosons 
divided by an equal number of IPC's. Normalization was obtained by requiring 
an equal content of simulated to experimental X/IPC events by 
multiplying the theoretical ratio with
a factor $u$, the transmission or inverse efficiency for boson emission. 
Boson masses were fitted from threshold to the transition energy in e.g 
64 channel bins using a ${\chi}^{2}$ procedure. 
The inverse of the ${\chi}^{2}$ was taken to be the transition probability: 
$P = ({\chi}^{2}/(N-1))^{-1}$, where $N$ is the number of detector combinations. 
The value $P$ can then be plotted as a function of assumed $m_{i}$,
testing the likelyhood of a particle with the given mass being present
in the spectrum. An upper limit for the branching ratio is obtained by multiplying $P$ with the 
'transmission' $u$.
For three detector geometries $10^{7}$ or $10^{8}$
events have been generated for some 40 to 70 boson masses, with and 
without energy disparity ($y$) cuts, for transition energies of 6.05, 8.87, 
10.96, 12.71
14.64, 17.23 and 17.64 MeV. Also for the IPC (E0, E1, E2, M1 and M2) 
distributions, generally up to $10^{9}$
events were simulated with and without $y$-cuts. 
It was found that a proportional scaling between 
boson masses and transition energies could be applied, except 
for the low masses, where
the transition from backward decay to forward two-body decay at ${\beta}_{CM} = 
{\beta}_{lab}$ is not scaled correspondingly. A detailed description  
of the program will be published elsewhere \cite{fokke5}.        

\begin{figure}
\includegraphics[height=.5\textheight]{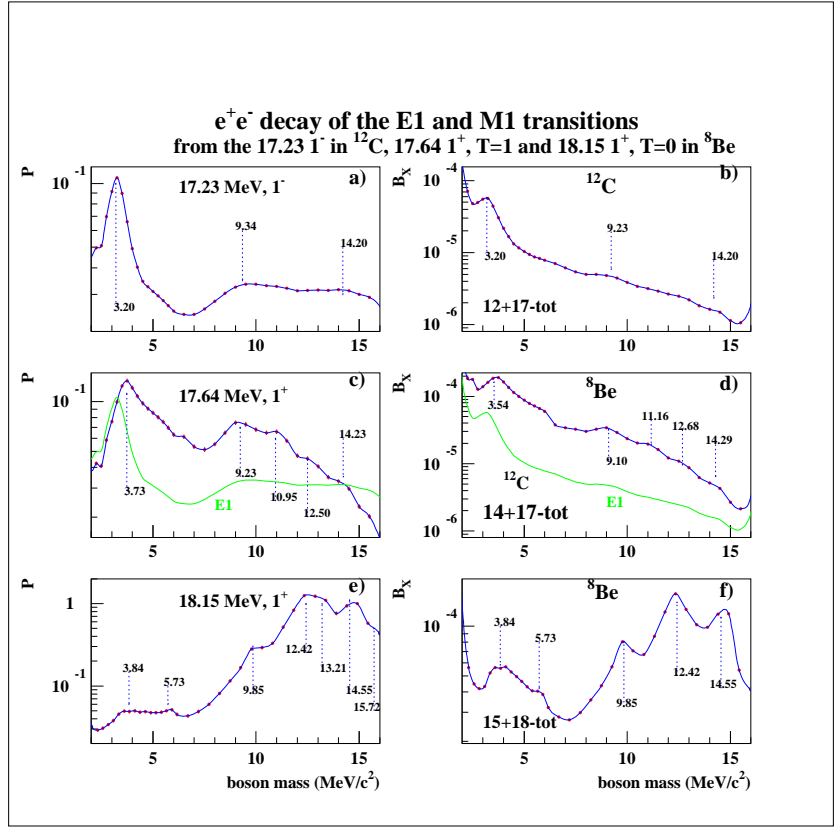}
\caption{ \small \em Figs.\, 1a, 1c and 1e show the probability distributions
of bosons in the decay of the 17,23 $1^{-}$, the 17.64 $1^{+}$ and the
the 18.15 MeV $1^{+}$ levels
following the $^{11}$B($p,e^+e^-)^{12}$C$^{*}$ reaction at
1.6 MeV and the $^{7}$Li($p,e^+e^-)^{16}$O$^{*}$ reactions at
 0.44 respectively 1.03 MeV as
measured at IKF in 1992 and 1998 \cite{fokke1,fokke4}. Figs.\, 1b, 1d and 1e
show the corresponding $B_X$-values i.e. branching ratios for boson to    
${\gamma}$-ray emission. The dashed lines in Figs.\, 1c and 1d correspond 
with the E1 distributions from Figs.\, 1a and 1b.}
\label{fig-bc1723}
\end{figure}
\subsubsection{The IKF experiments revisited}

To test the validity of the program we have first applied it to the earlier 
results at IKF. Spectra from the $^{7}$Li$(p,e^{+}e^{-})^{8}$Be$^{*}$ reaction
at 0.44 and 1.03 MeV and the $^{11}$B$(p,e^{+}e^{-})^{12}$C$^{*}$
reaction at 1.6 MeV proton energy were considered.
All 28 correlation angles were used in the fits.
The mass distributions perfectly confirm the 9 MeV/c$^{2}$ boson hypothesis
with the peak at 9.23 MeV in Fig.\,1c.
The second peak at 10.95 MeV corresponds with the emission of
the 9 MeV/c$^{2}$ boson in
the 14.64 MeV transition to the first excited state, being
at 9.09 MeV/c$^{2}$ in the 14.64 MeV laboratory system. No significant peak 
is seen at 9 MeV/c$^{2}$ in Fig.\,1a.
The $P$-value for $^{8}$Be at 9 MeV/c$^{2}$
is indeed a factor five larger than for $^{12}$C as observed in 
the angular distributions. However, we also see peaks at 
the higher masses of 12.50 and 14.23 MeV/c$^{2}$ masses.

A different picture is found for the isoscalar  
transition at 18.15 MeV. Here, the dominant peaks are at 12.42
and 14.55 MeV/$c^{2}$. They are a factor 20 stronger than in the isovector 
M1, whereas the strength of the 9.23 MeV/$c^{2}$ peak is about the same.
The masses in Fig.\,1 are indicated using the 
ground state transitions as two-body decay frame. Applying the 
transformation to 14.64, the masses become 2.99, 7.55, 9.26, 10.52 and 11.86 
respectively and in Fig.\,1f (with 15.15) of 3.21, 4.78, 8.22, 10.37 and
12.15 MeV. The emission of the same 9 MeV/c$^{2}$ boson  
in the 17.64 (9.10) and in the 14.64 MeV (9.23) was
already suggested in \cite{fokke2} where approximately  the same 
branching was deduced for this boson. 
Figs.\,1b, 1d and 1f show the branching ratios $B_X$ for boson to
${\gamma}$-ray emission. 
Integration over 9 MeV/c$^{2}$ yields respective values 
of $1.9{\cdot}10^{-5}$, $1.2{\cdot}10^{-4}$ and $4.0{\cdot}10^{-4}$
within 1${\sigma}$ in accord with the values of 
${\leq}2.3{\cdot}10^{-5}$, $(1.1{\pm}0.3){\cdot}10^{-4}$ and $(5.8{\pm}2.2){\cdot}10^{-4}$ given in \cite{fokke1,fokke2,fokke3,fokke4}. 
The mass distributions for the transitions after
applying symmetry cuts to the spectra. 
show dramatically larger  variations than in Fig.\,1 
at 9.0, 12.5 and 14.5 MeV/c$^{2}$, in agreement with the expectations
for two-body decay where the symmetrical decay is enhanced with
respect to the primarily asymmetrical decay of M1-IPC at large angles. 
Integration at 9 MeV/c$^{2}$ yields 
$B_{X}$ values of about half of those
obtained from Fig.\,1, in agreement with the expectations for two-body decay.   

It appears that the boson search program adequately describes
the IKF-results by reproducing the previously found signal 
at 9 MeV/c$^{2}$. 
However, surprisingly, it also finds signals for particles
at the masses of 12.5 MeV/c$^{2}$ and 14.5 MeV/c$^{2}$. 
In the calibration runs: $^{19}$F$(p,{\alpha}e^{+}e^{-})^{16}$O$^{*}$  
at 0.72 and 0.84 MeV just below and above threshold for the 8.87 MeV $2^{-}$ 
level, two boson candidates are observed at 5.0 and 5.38 MeV/c$^{2}$ with
branching ratios as large as several percent. 

\subsubsection{Back to ATOMKI and further}

Because of the apparent feasibility of transformation of the 
angular spectra to mass-spectra and the unbiased character
of the program we have systematically converted all available angular 
spectra from IKF and from ATOMKI and other sources into mass spectra. 
The oldest investigated spectrum was
the celebrated angular correlation measurement of the E0 transition at 6.05
MeV in $^{16}$O dated back to 1949 \cite{lindsey} and the most recent one:
the study of the $e^{+}e^{-}$ decay of the giant resonance decay at 22.23 MeV 
and the 15.11 MeV isovector M1 decay in $^{12}$C at Stony Brook  
\cite{hatzi} in 2003.

\begin{figure}
\includegraphics[height=.625\textheight]{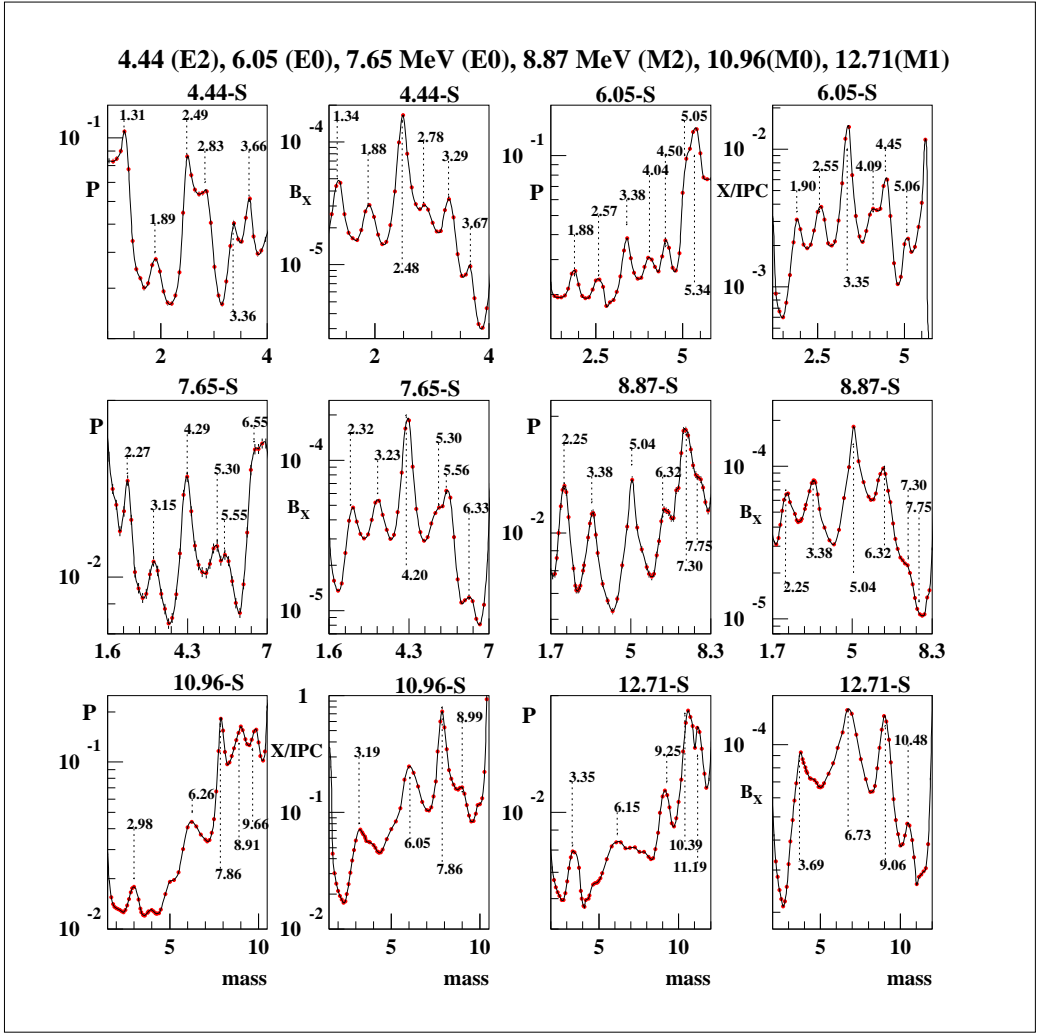}
\caption{\small \em From top to bottom (left) are shown the probability distributions 
of symmetrical (S) pairs in the 4.44 (E2), 6.05 (E0), 7.65 (E0) 
and 8.87 (M2), 10.96 (M0) and 12.71 (M1) transitions in $^{12}$C and $^{16}$O. 
Right are shown the corresponding branching ratios for boson emission
with respect to E0-IPC $(X/IPC)$ for the 6.05 MeV E0 and the 10.96 MeV (M0) 
transitions and to ${\gamma}$-rays for the E2 at 4.44 MeV, the M2 at 8.87 and 
the M1 at 12.71  MeV. The $B_X$-value for the 7.65 MeV E0 is shown with respect to the parallel $0^{+}{\rightarrow}2^{+}$ 3.20 MeV E2-${\gamma}$-ray.}
\label{fig-bigmacs2}
\end{figure}

\subsubsection{The 4.44, 6.05, 7.66, 8.87, 10.96 and 12.71 MeV transitions} 

In this paragraph, we discuss the mass-distributions of six different
transitions in $^{12}$C and $^{16}$O, selected from results of the  
$^{12}$C$(p,p'e^{+}e^{-})^{12}$C$^{*}$ and the $^{16}$O$(p,p'e^{+}e^{-})^{16}$O$^{*}$ reactions at 16.5 MeV and from the 
$^{19}$F$(p,{\alpha}e^{+}e^{-})^{16}$O$^{*}$ reaction used for calibration. 

The mass-distributions for symmetrical (S) pairs are shown in Fig.\,2, with 
the $P$-distribution in the left column and the branching ratio in the right. 
In all shown data, disparity cuts of $y {\leq} 0.5$ have been applied.
For every transition more than one spectrum is available showing 
in essence the same features as Fig.\,2. In particular, for the 6.05 MeV E0 
transition, data from twelve experiments have been analyzed. Eight of them 
show strong peaks at 5.0 and 5.3 MeV/c$^{2}$. The four spectra where 
they are absent, stem from experiments where the maximum correlation angle
is smaller than 110$^{0}$.
 
The data are discussed from bottom to top. right to left. In the 12.71 MeV transition,
we clearly recognize the boson at 9 MeV/c$^{2}$, the original goal for the experiments in Debrecen. The branching ratio is about $8{\cdot}10^{-4}$ 
in agreement with the results of \cite{fokke2,fokke3,fokke4}.
However, there are two new candidates at 10.39 and 11.19 MeV/c$^{2}$, 
which are stronger than the one at 9 MeV/c$^{2}$. 
The M1 data are compared (bottom left) with a measurement of the
angular distribution of the {\em forbidden} $0^{+} {\rightarrow} 0^{+}$
M0 transition at 10.96 MeV in $^{16}$O, where a strong signal is 
seen at 7.85 MeV/c$^{2}$.  
The X/IPC value is as large as 0.8 and for the ($y{\leq}1$) cut (not shown here)
it is even 2, suggesting that the decay via the 7.85 MeV/c$^{2}$ boson is the 
dominant decay mode in this M0 transition. The two peaks at 8.99 and 9.84 MeV 
can be ascribed to reflection from bosons of 7.35 and 7.85 MeV/c$^{2}$ both 
occurring in the decay of the neighboring 8.87 MeV/c$^{2}$ which could 
experimentally not be completely resolved from the 10.96 MeV sum-peak.   
It seems appropriate to assign, from the superallowed decay, 
the 7.85 MeV/c$^{2}$
as a pseudoscalar particle and for the same reason the signals at 9.23
and 10.39 MeV/c$^{2}$ as axial vector particles. A vector character would be 
preferred for the 11.19 MeV/c$^{2}$ signal. 

In the 8.87 MeV M2 transition we see a boson candidate at 7.35 with the 
spin $1^{-}$ and possibly the 7.85 $0^{-}$. 
The 6.32 MeV/c$^{2}$ peak could be due to reflection of 
the 7.85 in the 10.96 transition. However, the signals at 2.25. 3.38 and 5.04 
MeV/c$^{2}$ in this transition are all viable boson candidates.   
Also in the mass-distributions of the  
two E0 transitions at 6.05 and at 7.65 MeV,  
multiple resonances are displayed. A candidate boson at 3.36 
MeV/c$^{2}$ is seen in all spectra,
The occurrence in the E0, E2 and M2 transitions (allowed decay) could only be 
compatible with the spin-parity of $1^{-}$ for this particle. With the 
simultaneous presence of the candidates below 4 MeV in the E0 and E1 
transitions   
the spin is restricted to $1^{-}$. Likewise, the resonance at 5 MeV/c$^{2}$ 
must, from its presence in the E0's and the M2 transition, have spin-parity 
$1^{-}$. 
With its appearance in both E0 transitions and its absence in the M2,
the 4.48 MeV/c$^{2}$ candidate must have a scalar $0^{+}$ character since a 
$1^{-}$ particle would also have shown up in the M2 transition. With some 
optimism, this figure
indicates the existence of two scalar $X$-bosons at 4.5 and 5.3 MeV/c$^{2}$
and six vector bosons at 2.5, 3.3, 4.0, 5.0, 6.3 and 7.3 MeV/c$^{2}$.

\subsection{Conclusion and the future}

It appears that in every isoscalar transition considered with energies
from 4.44 to 12.71 MeV deviations from IPC are observed.
This anomaly is explained in an isoscalar particle scenario 
which would require a multitude of bosons. The finding could 
have the bizarre consequence of a new type of light neutral matter 
instead of or/and the discovery of a new gauge boson. 

A second generation of dedicated experiments is 
required to test this exciting possibility.
The properties of every boson candidate have to be checked in great detail. 
A first experiment in that direction \cite{fokke5} has been 
undertaken in the DEB7 run, where the two-body decay properties of the 
bosons were verified, by mounting the central
detector combination at an asymmetrical position.
An alternative approach is being performed \cite{kraszna1} at ATOMKI to 
study the angular correlations with an order of magnitude higher angularity, 
using wire chambers in front of the $dE/dx$-detectors. 

\subsection{Acknowledgement}

This paper has been presented at the International Symposium on Exotic Nuclear
Systems, ENS'05, at ATOMKI, Debrecen, Hungary, 20-25 June 2005.

I am indebted to my colleagues: J. van Klinken (KVI, Groningen), H. Bokemeyer (GSI, Darmstadt), K. Bethge, O. Fr\"{o}hlich, K.A. M\"{u}ller,  
K.E. Stiebing (IKF, Frankfurt), R. van Dantzig,  T.J.Ketel (NIKHEF, Amsterdam),  K. Griffioen (William and Mary, Williamsburg), A. van der Schaaf (Universit\"{a}t Z\"{u}rich, Z\"{u}rich), 
M. Csatl\'{o}s, L. Csige, Z. G\'{a}czi, J. Guly\'{a}s, M. Hunyadi, A. Vit\'{e}z, A. Krasznahorkay Jr. and A. Krasznahorkay (ATOMKI, Debrecen).


{\small
\begin{thebibliography}{aipproc}


\bibitem{rose} M.E. Rose, Phys. Rev. 76 (1949) 678.
\bibitem{stiebing} K.E. Stiebing, F.W.N. de Boer, O. Fr\"{o}hlich, H. Bokemeyer,
K. A. M\"{u}ller,\\ K. Bethge and J. van Klinken,
J. Phys. G: Nucl. Part. Phys. 30 (2004) 165. 
\bibitem{fokke1} F.W.N. de Boer, O. Fr\"{o}hlich, K.E. Stiebing, K. Bethge,
H. Bokemeyer, A. Balanda,\\ A. Buda, R. van Dantzig, Th. W. Elze, H. Folger, 
J. van Klinken, K. A. M\"{u}ller,\\ K. Stelzer, P. Thee and M. Waldschmidt,
Phys. Lett B388 (1996) 235.
\bibitem{fokke2} F.W.N. de Boer, R. van Dantzig, J. van Klinken, K. Bethge,
H. Bokemeyer, A. Buda, \\K. A. M\"{u}ller and K.E. Stiebing, 
J. Phys. G: Nucl. Part. Phys. 23 (1997) L85.
\bibitem{fokke3} F.W.N. de Boer et al., Nucl. Phys. B72 (1999) 189.
\bibitem{fokke2} F.W.N. de Boer et al., Nucl. Phys. B72 (1999) 189.
\bibitem{fokke4} F.W.N. de Boer et al. J. Phys. G: Nucl. Part. Phys. 27 (2001) L29, 
\bibitem{bromley} D.A. Bromley et al., Phys. Rev. 114 (1959) 758. 
\bibitem{almquist} E. Almquist et al., Phys. Rev. 114 (1959) 1040. 
\bibitem{lindsey} S. Devons and G.R. Lindsey, Nature 164 (1949) 539 and \\
S. Devons and G. Goldring, Proc. Phys. Soc. A67 (1955) 131. 
\bibitem{hatzi} A. Hatzikoutelis, PhD-Thesis, 2003, Stony Brook University.
\bibitem{fokke0} F.W.N. de Boer, J. Deutsch et al., J. Phys. G: Nucl. Part. 
Phys. 14 (1988) L131. 
\bibitem{fokke5} F.W.N. de Boer et al., to be publ. in J. Phys. G: Nucl. Part.
Phys. Letter to the Editor.
\bibitem{kraszna1} A. Krasznahorkay, F.W.N. de Boer, J. Guly\'{a}s, Z. G\'{a}czi,
T.J. Ketel, M Csatl\'{a}s, L. Csige,\\ M. Hunyadi, J. van Klinken, A. Krasznahorkay Jr.,
A Vit\'{e}s, to be publ. in Phys. Rev. Lett.  
\end{thebibliography}

}
\end{document}